\documentclass[usenatbib]{mnras}

\usepackage{graphicx}	
\usepackage{amsmath}	
\usepackage{amssymb}	

	\title[Powerful feedback in the LLAGN ESO\,428-G14]{Powerful mechanical-driven outflows in the central parsecs of the Low-Luminosity Active Galactic Nucleus ESO\,428-G14}
	    \author[D. May et al.]{
    D. May $^{1,2}$\thanks{E-mail: dmay.astro@gmail.br},
    A. Rodr\'iguez-Ardila $^{2,3}$,
		M.A. Prieto $^{3,4}$,
		J.A. Fern\'andez-Ontiveros $^{3,4}$,
		\newauthor
		Y. Diaz $^{5,6}$,
		X. Mazzalay $^{7,8}$
    \\
    	$^{1}$Instituto de Astronomia, Geof\'isica e Ci\^encias Atmosf\'ericas, Universidade de S\~ao Paulo, 05508-090, S\~ao Paulo, SP, Brazil
		\\
		$^{2}$Laborat\'orio Nacional de Astrof\'isica/MCTI, 37530-000, Itajub\'a, MG, Brazil
		\\
		$^{3}$Instituto de Astrof\'isica de Canarias, C/V\'ia L\'actea s/n, E-38205, La Laguna, Tenerife, Spain
		\\
		$^{4}$Universidad de La Laguna, Dpto. Astrof\'isica, Avd. Astrof\'isico Fco. S\'anchez s/n, E-38206, La Laguna, Tenerife, Spain
		\\
		$^{5}$Instituto Nacional de Pesquisas Espaciais/MCTI, 12227-010, S\~ao Jos\'e dos Campos, SP, Brazil
		\\
		$^{6}$Instituto de F\'isica y Astronom\'ia, Facultad de Ciencias, Universidad de Valpara\'iso, Gran Bretana No. 1111, Playa Ancha, Valpara\'iso, Chile
		\\
		$^{7}$Max-Planck-Institut f\"ur extraterrestrische Physik, Postfach 1312, D-85741, Garching, Germany
		\\
		$^{8}$Universit\"ats-Sternwarte Ludwig-Maximilians-Universit\"at, Scheinerstrasse 1, D-81679 M\"unchen, Germany}
	
\begin{document}

    \date{Draft for internal use only}

    \pagerange{\pageref{firstpage}--\pageref{lastpage}} \pubyear{2018}

    \maketitle

    \label{firstpage}

    \begin{abstract}
    
    Low-Luminosity Active Galactic Nuclei (LLAGN) are characterized for low radiative efficiency, much less than one percent of their Eddington limit. Nevertheless, their main energy release may be mechanical, opposite to powerful AGN classes like Seyfert and Quasars.
This work reports on the jet-driven mechanical energy and the corresponding mass outflow deposited by the jet in the central 170~parsecs of the nearby LLAGN ESO\,428-G14. The jet kinetic output is traced through the coronal line [Si\,{\sc vi}] $\lambda$19641 \AA. It is shown that its radial extension, up to hundreds of parsecs, requires a combination of photoionization by the central source and shock excitation as its origin.
 From the energetics of the ionized gas it is found that the mass outflow rate of the coronal gas is in the range from 3$-8$~M$\odot$~yr$^{-1}$, comparable to those estimated from H{\sc i} gas at kiloparsec scales in powerful radio  galaxies.

    \end{abstract}


    \section{Introduction}
    \label{sec:intro}
    
    Low-Luminosity Active Galactic Nuclei (LLAGN) are characterized by luminosities $\leq 10^{42}$\,erg~s$^{-1}$, comparable to that of starburst galaxies. Most of their light is dominated by that of their host, often associated to early type galaxies. The low radiative efficiency of LLAGN is at odds with the often large supermassive black hole lurking at their nuclei, which place them in the low Eddington regime of accreting black holes ($L_{\rm bol}/L_{\rm Edd} < 10^{-3}$). LLAGN often present jets and thus a main channel for releasing energy may be via mechanical feedback \citep{Merloni12}, opposite  to classical powerful AGN, where the radiative channel is the dominant one. 
    
    A direct estimate of jet-driven shocks would require a kinematic tracer that is spatially associated to the jet passage through the ISM of the galaxy. \citet{Morganti05} use the 21~cm absorption to measure energetic ($\sim$1000~km\,s$^{-1}$) massive outflows (1.2$-$50~M$\odot$\,yr$^{-1}$) at kiloparsec distances from the core of powerful radio galaxies, most likely driven by such interactions. 
    
    Here, we report on comparable massive outflows traced by the high ionization [Si\,{\sc vi}] $\lambda$19641 \AA~emission line but at a few hundred parsecs from the core of the LLAGN ESO\,428-G14, a nearby ($z=$ 0.00566; D=19.5~Mpc; 1"= 92 pc) Seyfert\,2 SAB galaxy, with $L_{\rm bol}=2\pm1\times$10$^{42}$ erg s$^{-1}$ (this work, Sect.~\ref{sec:shocks}). The galaxy has a two-side bent jet spatially correlated with [O\,{\sc iii}], H$\alpha$+[N\,{\sc ii}] and X-rays \citep{Falcke96,Fabbiano17}. Convincing observational evidence of strong feedback in LLAGN are scarce. Thanks to the excellent angular resolution, we map the kinematics of the emission line gas that interacts with the radio jet by means of a unique AGN tracer. This makes ESO\,428-G14 the most rich showcase of outflows in a LLAGN studied so far.

    \section{Observations and data reduction}
    \label{sec:2}

    We obtained adaptive-optics-assisted $K-$band integral field spectroscopy with SINFONI \citep{Eisenhauer03} at the Very Large Telescope (VLT). The observations were performed under programme 086.B-0484(A) and were taken on the nights of December 22, 2010 and January 11, 2011, with an individual exposure time of 300~s and a total integration of 2700~s on-source. The pixel scale provided by the detector is 0.05 arcsec~$\times$~0.1 arcsec. The spectra range from 1.95 to 2.45 $\mu$m, with a corresponding spectral resolution of v$\sim$60 km s$^{-1}$. The reduction routines were completed using the EsoReflex pipeline. 
    Two sets of 3 data cubes each were combined to obtain the extended emission-line maps, with a FoV of 3.35 arcsec~$\times$~2.62 arcsec (hereafter data set 1 - DS1) and a second data cube with high signal-to-noise (S/N) spectra was obtained combining 9 observations (FoV of 1.85 arcsec~$\times$~1.75 arcsec - hereafter DS2). The resulting angular scale is $\sim$0.16~arcsecs.
    After the reduction, we performed the data treatment procedure described in \citet{Menezes15}.
    
    In addition, long-slit optical spectra were obtained using the Goodman spectrograph \citep{CCA04} attached to the 4.1~m SOAR Telescope atop Cerro Pach\'on, Chile. The spectra were collected in March 10, 2017 using a 400 l/mm grating and a 0.8 arcsec slit width at a position angle of 135$^{\circ}$ ~East of North,  following the extended [\ion{Si}{vi}] emission and the radio-jet. They cover the wavelength interval of 3370$-$7380~\AA\ with R$\sim$1000 at 5500~\AA. 
   The spectra were reduced following standard IRAF procedures. Three extractions, one centred in the nucleus (1.6 arcsec size) and two additional ones (1 arcsec size) at both sides of the nucleus were made. These are meant to be co-spatial to the SINFONI data. However, a simple point-to-point comparison between the optical and near-infrared (NIR) observations cannot be made because of the seeing-limited nature ($\sim$1") of the Goodman spectroscopy.

   \section{Results}
		\subsection{The coronal line emission}
    \label{sec:si}

		\begin{figure*}
    \resizebox{\hsize}{!}{\includegraphics{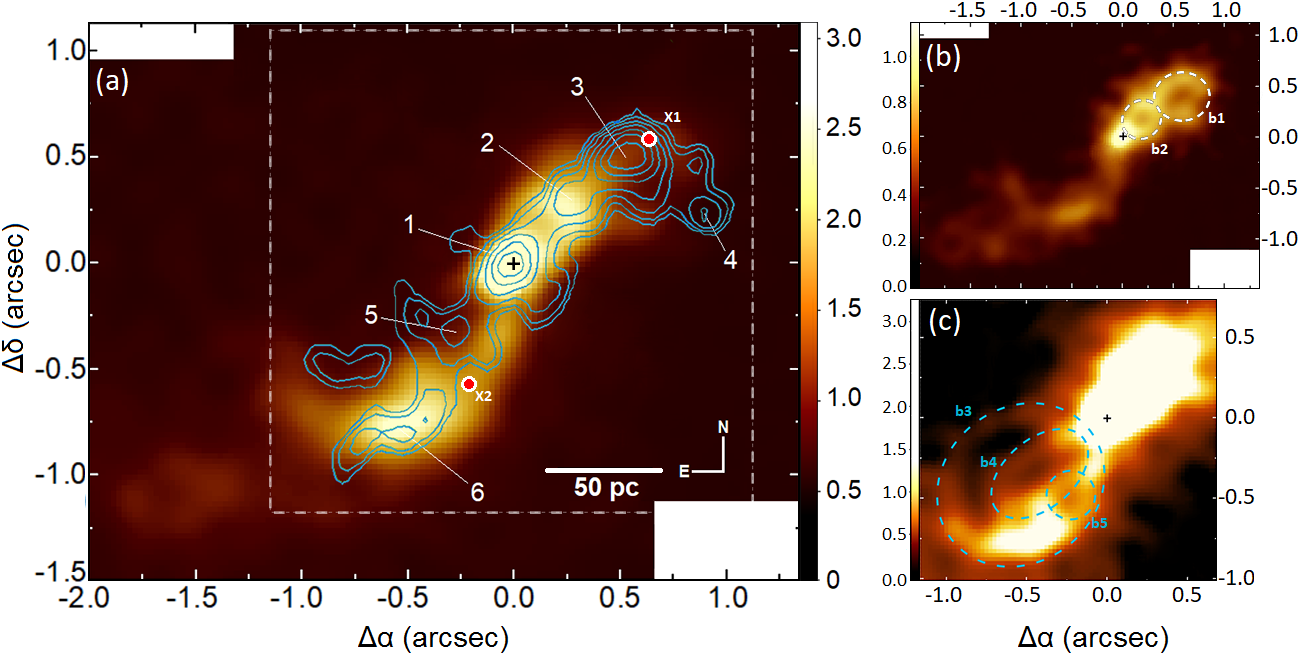}}
    \caption{Panel (a): mosaic of two sets of data cubes (DS1) for the [Si\,{\sc vi}] $\lambda$19641 \AA~emission line (total flux of $58\pm3\times 10^{-15}$~erg s$^{-1}$~cm$^{-2}$) overlaid with the VLA 2\,cm radio emission (blue contours); the dashed square denotes the VLA FoV and the red dots x1 and x2 the extended \textit{Chandra} X-ray peaks. (b): Image of the H\,{\sc i}\,Br$\gamma$ $\lambda$21661 \AA~emission (total flux of $24\pm1\times 10^{-15}$~erg s$^{-1}$~cm$^{-2}$); the dashed ellipses highlight the ``helix'' in two substructures, b1 and b2. (c): The [Si\,{\sc vi}] line from all the combined data cubes (DS2), with smaller FoV and higher signal-to-noise ratio. The enhanced contrast shows the faintest structures b3, b4 and b5. The cross marks the AGN position.The flux bar is in units of $10^{-19}$~erg s$^{-1}$~cm$^{-2}$~\AA$^{-1}$.}
     \label{fig:si}
    \end{figure*}	
				
		With an ionization potential (IP) of 168\,eV, the [Si\,{\sc vi}] $\lambda$19641\AA~emission is one of the best tracers of highly ionized gas in the NIR, directly associated to the most energetic excitation mechanisms in AGNs \citep{Marconi94}. Fig.~\ref{fig:si}a shows the emission map for the [Si\,{\sc vi}] line. The gas morphology is extended, reaching up to 195\,pc to the SE and 110~pc to the NW. 
		Up to our knowledge, no other LLAGN has been reported in the literature with such a remarkable coronal line (CL) emission. The coronal gas is distributed along a PA very close to that of the radio jet, forming multiple filaments and knots of emission, similar to the [\ion{Fe}{ii}] gas morphology reported by \citet[][hereafter RSWB06]{RSWB06}. We defined the AGN location as the position where the $K-$band continuum emission (2.2~$\mu$m) peaks, which coincides with the maximum intensity of the [Si\,{\sc vi}] emission. Secondary peaks of emission are also observed, being the most prominent ones those located at $\sim$85~pc SE (region 6) and at $\sim$35~pc NW from the centre (region 2). The former consists of a conspicuous bright spot while the later is associated to a very peculiar structure, consisting of thin strands of ionized gas following an helical-shape. These are best seen in the Br$\gamma$ image shown in panel~\ref{fig:si}b and also described in the optical by \citet{Falcke96}. We will refer to this region as "helix", comprising two substructures b1 and b2.
		When the brighter regions are saturated in the scale palette, fainter structures become evident in DS2, as shown in panel~\ref{fig:si}c. Region 6, for instance, is part of a larger ring-like feature linked to the nucleus through a faint short filament of gas. We marked this external ring as b3. We also detected two fainter inner rings within b3 (b4 and b5). We use the major axis of b3 to define the PA of the ionization cones, resulting in 130$^{\circ}$$\pm$3$^{\circ}$.
		
		 \citet{prieto05} reported imaging of the isolated coronal line [\ion{Si}{vii}]~24830~\AA\ (IP=206~eV), done with the adaptive optics assisted IR camera NACO at the ESO/VLT. They found that ESO~428-G14 had the largest and best-collimated [\ion{Si}{vii}] emission of their sample, of up to 150-pc in radius from the nucleus (see their Figure~2), showing an excellent agreement with the [Si\,{\sc vi}] emission.

	\subsection{Kinematics of the coronal gas}
    \label{sec:sik}	
    
\begin{figure}
    \resizebox{\hsize}{!}{\includegraphics{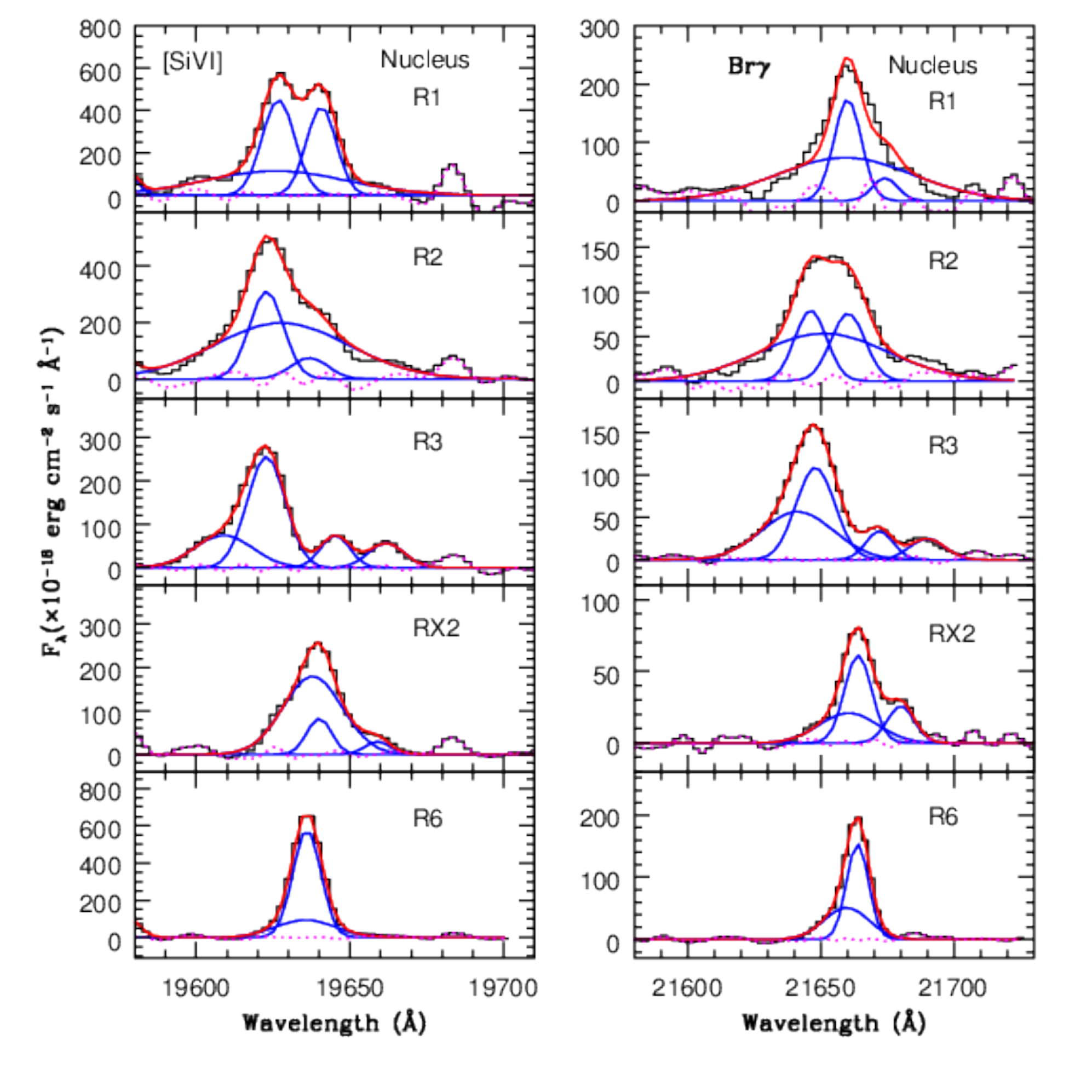}}
    \caption{Deblending of the [Si\,{\sc vi}]~19641~\AA~ line
    observed at the nucleus, region~2 (35~pc NW), region~3 (70~pc NW), region X2 (60~pc SE) and region~6 (85~pc SE). In all cases, the spectra were extracted using an aperture of 0.4"$\times$0.4" in size. The black histogram is the observed spectrum, the blue lines mark the individual Gaussian components, the red line is the total fit and the dotted magenta line is the residual after subtracting the total fit.}
    \label{fig:linefit}
    \end{figure}

     ESO\,428-G14 reveals a very complex velocity field  in the ionized gas, unveiled by the high angular resolution of the data. At most locations, the lines display a double peaked structure with broad wings.
    We deblended the observed profiles at selected positions of the FoV into individual components using LINER, a $\chi$-square minimization algorithm that fits up to eight different Gaussians to a line or set of blended lines. Figure~\ref{fig:linefit} shows the profiles of [Si\,{\sc vi}] and Br$\gamma$  at positions 1, 2, 3, x2 and 6 (Fig.~\ref{fig:si}a). In all cases the spectra are extracted with an aperture of 0.4$\arcsec \times 0.4\arcsec$ in size.
    
     Typically, three Gaussian components were necessary to model the profiles: two narrow components (FWHM$\leqslant$200~km\,s$^{-1}$, labeled as NC in Table~\ref{tab:linefit}) on top of an underlying broader line (BC). Indeed, this latter feature, with FWHM varying from 350~km\,s$^{-1}$ and up to $\sim$830~km\,s$^{-1}$, prevails across most of the regions where the ionized gas is detected. The consistency of the fit for both [Si\,{\sc vi}] and Br$\gamma$ in terms of the number of components and the parameters of the individual Gaussians allow us to conclude that they represent distinct kinematic system in the NLR. Table~\ref{tab:linefit} lists the parameters of each component (flux and FWHM) as well as the corresponding observed line flux ratio [Si\,{\sc vi}]/Br$\gamma$. The FWHM have been corrected for instrumental broadening.

    \begin{table}
    \begin{center}
    \caption[Line Fit]
    {Gaussian fits to [\ion{Si}{vi}] and Br$\gamma$ in ESO\,428-G14.}
    \tabcolsep=0.11cm
    \begin{tabular}{lcccccc}
    \hline \hline
Region$^{1}$ &  [\ion{Si}{vi}]$^{2}$  &  FWHM$^{3}$  &  Br$\gamma^{2}$ & FWHM$^{3}$ & [\ion{Si}{vi}]/Br$\gamma$ \\
\hline
1-NC1  & 6.0$\pm$0.5     & 181 & 2.2$\pm$0.1     &	155	 &	2.7$\pm$0.3  \\
1-NC2  & 5.1$\pm$0.5	 & 163	 & 0.5$\pm$0.1   &	140	 &	11.2$\pm$3.5 \\
1-BC1  & 6.6$\pm$2.2	 & 830	 & 4.7$\pm$0.7   &	828	 &	1.4$\pm$0.5 \\
2-NC1  & 4.3$\pm$0.3	 & 189	 & 1.2$\pm$0.1   &	184	 &	3.6$\pm$0.3 \\
2-NC2  & 1.1$\pm$0.3	 & 205	 & 1.2$\pm$0.1   &	199	 &	0.9$\pm$0.2 \\
2-BC1  & 10.8$\pm$1.0	 & 783	 & 3.2$\pm$0.3   &	787	 &	3.3$\pm$0.4 \\
3-NC1  & 4.0$\pm$0.1	 & 213	 & 2.0$\pm$0.1   &	228	 &	2.0$\pm$0.1 \\
3-NC2  & 0.9$\pm$0.1	 & 160	 & 0.4$\pm$0.1   &	155	 &	2.1$\pm$0.4 \\
3-NC3  & 0.8$\pm$0.1	 & 204	 & 0.4$\pm$0.1   &	205	 &	2.1$\pm$0.6 \\
3-BC1  & 1.8$\pm$0.2	 & 338	 & 1.9$\pm$0.2   &	439	 &	0.9$\pm$0.1 \\
6-NC1  & 6.6$\pm$0.1	 & 152	 & 1.6$\pm$0.1   &	120	 &	4.2$\pm$0.1 \\
6-BC1  & 2.3$\pm$0.3	 & 345	 & 1.1$\pm$0.1   &	270	 &	2.1$\pm$0.3 \\
\hline
\end{tabular}
\begin{minipage}{8cm}
$^1$Each region is identified according to the numbers given in Figure~\ref{fig:si}a. NC and BC stand for narrow component and broad component, respectively; $^2$ Flux in units of 10$^{-15}$~erg\,cm$^{-2}$\,s$^{-1}$; ~  $^3$Units of km\,s$^{-1}$
\end{minipage}
\label{tab:linefit}
\end{center}
\end{table}

\begin{figure*}
    \resizebox{\hsize}{!}{\includegraphics{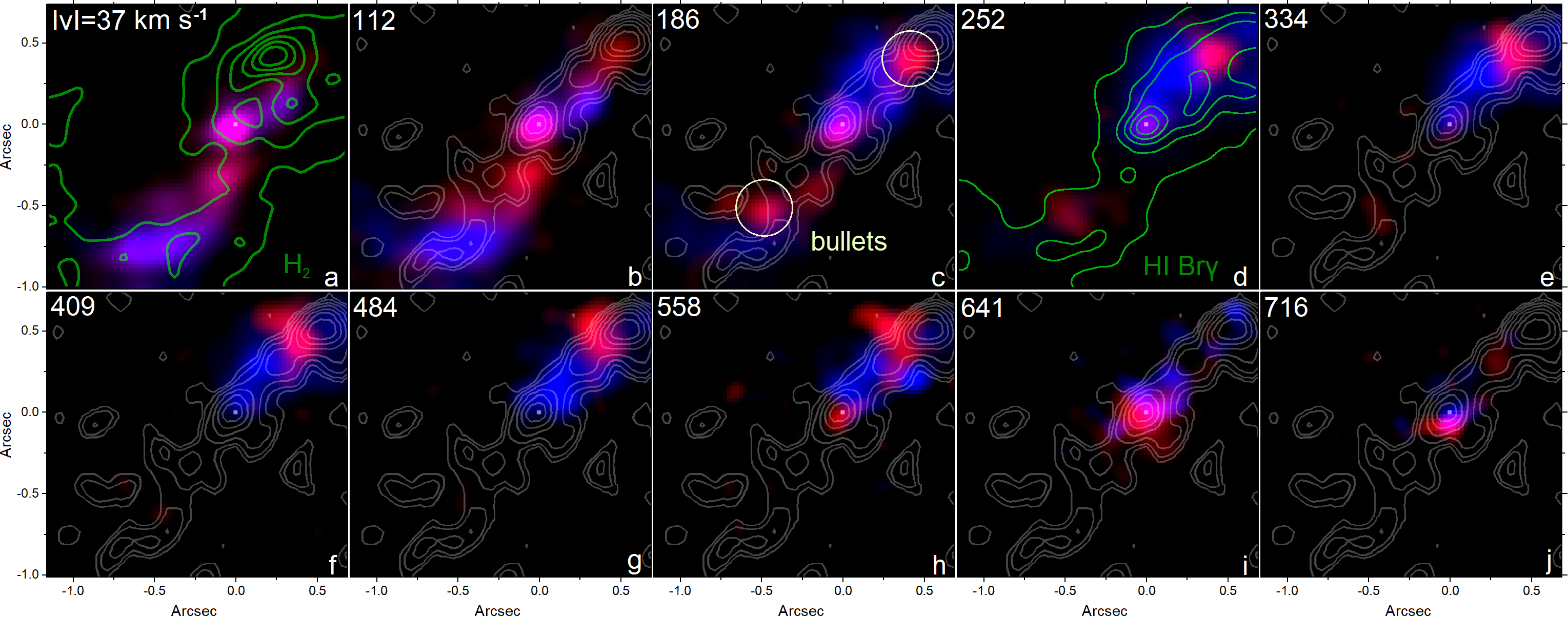}}
    \caption{BRV maps (blueshifted and redshifted velocity maps) for the [Si\,{\sc vi}] $\lambda$19641 \AA~emission line. The velocity ranges from $|v|\lesssim$700 km s$^{-1}$, with steps of $\sim$75 km s$^{-1}$ and the corresponding absolute velocity of each map is shown at the top. Contours denote the radio (white), the H$_{2}$ $\lambda$21218 \AA~(panel a) and Br$\gamma$ $\lambda$21661 \AA~(panel d) emissions. The cross marks the AGN centre.}
     \label{fig:sikin}
    \end{figure*}

    In Fig.~\ref{fig:sikin} we show BRV maps (blueshifted and redshifted velocity maps), which depicts in blue and red colours the gas in blueshift and redshift, respectively, with nearly the same projected velocities.  
    The most striking results are: \textit{(i)} the [Si\,{\sc vi}] gas at the nucleus and to the NW spans a large velocity interval ($\mid v\mid\lesssim$700 km\,s$^{-1}$), nearly twice the value reported by RSWB06 for [\ion{Fe}{ii}]~12570~\AA.  These high-velocity clouds fully agree with the detection of broad wings in the ionized gas shown in Fig.~\ref{fig:linefit}; \textit{(ii)} no clear signs of rotation, with blueshifted and redshifted velocity values found in both sides of the cones; \textit{(iii)} region 6 is dominated by low-velocity gas, with $\mid v\mid<$200 km\,s$^{-1}$ (panels a, b and c of Fig.~\ref{fig:sikin}); \textit{(iv)} the detection of two [Si\,{\sc vi}] redshifted ``bullets'' at opposite directions from the AGN and aligned with the radio-jet axis (panel c). The SE bullet is located at 0.6 arcsec ($\sim$55\,pc) from the AGN, next to region 6, while the NW counterpart (at 0.7 arcsec, or $\sim$65\,pc) is half-way between regions 2 and 3 in Fig.~\ref{fig:si}a.

	\subsection{Comparison to the radio and X-ray emissions}
    \label{sec:radio}
    
    In Fig.~\ref{fig:si}a we overlapped the 2\,cm radio map from VLA \citep{Falcke98} to the [Si\,{\sc vi}] emissions by matching both the radio and [Si\,{\sc vi}] brightest peaks. Taking into account regions 1, 2, 3 and 5, the orientation of the radio emission is 135$^{\circ}$ $\pm$ 2$^{\circ}$, similar to that of [\ion{Si}{vi}]. In addition to the nucleus, regions 2 and 6 have their peaks roughly coincident in radio and [Si\,{\sc vi}]. In the NW side, the jet bends abruptly $\sim$90$^{\circ}$~(region 3) to the SW (region 4), where it seems to fill the gap/void between two strands of gas. All these features suggest a strong association between [Si\,{\sc vi}] and the radio jet, confirming previous results by \citet{prieto05} and RSWB06 using other NIR lines.
    
    X-ray observations obtained with \textit{Chandra} show hard X-ray and Fe\,K$\alpha$ emission extending up to 2~kpc to the SE \citep{Fabbiano17}. It has a PA similar to that of [Si\,{\sc vi}]. Moreover, three peaks of X-ray emission are found: one at the nucleus and two others at regions x1 and x2 shown in Fig.~\ref{fig:si}a. Region x1 lies in the vicinity where the jet is bent. Note also that regions x1 and x2 are not co-linear with the AGN and peak near to radio knots 3 and 5, respectively, with x2 centred in b5. The X-ray emission also seems to be systematically displaced to the SW side along the [Si\,{\sc vi}] structure b3.
     Recently, \citet{fabbiano18} carried out a detailed spectral analysis of the nuclear and extended X-ray emission of ESO~428-G14, showing that shocks, of at least a few hundred km~s$^{-1}$ are present in the central, 170 pc (1.5") radius.  

		\section{Discussion}
    \label{sec:discussion}
		
		\subsection{The nature of ionized extended emission}
		\label{sec:shocks}

		 The simultaneous detection of extended CL emission, up to distances of 170~pc from the AGN, broad components in the ionized gas $-$not easily ascribed to rotation around a supermassive black hole$-$ and kiloparsec-scale X-ray emission, support earlier claims that mechanical heating should be at work in this object \citep{prieto05}. Shocks, triggered by the interaction between the radio-jet and the ISM, in addition to photoionization by the AGN, should enhanced the observed extended coronal emission. An immediate prediction for the presence of shocks is the production of free-free emission, with a maximum in the UV to X-ray from the shocked-heated gas, as already was confirmed. Moreover, \citet{prieto05} estimated that shocks with velocities of up to 900~km\,s$^{-1}$ would be able to explain the extended X-ray morphology. The lack of suitable data at that time hindered to fully test this scenario.

	We investigate if photoionization by the central source alone is able to sustain highly ionized clouds at the distances detected in ESO\,428-G14. To this aim we run grids of photoionization models with CLOUDY (version C17.01, \citealt{Ferland2017}). The input parameters include the gas density, $n_{\rm 0}$; the distance of the clouds to the nucleus, $R$; the spectral energy distribution of the ionizing radiation ("Table AGN" command, similar to that deduced by \citealt{Mathews87}), solar abundances and clouds with no dust.  We estimated a total AGN luminosity of 2$\pm1 \times 10^{42}$ erg\,s$^{-1}$. This value was obtained after direct integration of mid- to near-IR SED, using the VLT/VISIR fluxes given by \citet{Asmus14}  and the NIR flux at $2.12\,\mu$m from our VLT/NACO observations \citep{prieto14}, once corrected by a foreground Galactic extinction of $A_V = 0.54\, \rm{mag}$. The absorption corrected X-ray luminosity is negligible when compared to the IR contribution($\lesssim 10^{41}$~erg\,s$^{-1}$).
	
     The optical spectra allow us to set a lower limit to $n_0$ because of the detection of forbidden lines that are sensitive to the gas density.
      In order to determine $n_0$ we employed equation~5 of \citet{Proxauf14}. A gas temperature of $T_{\rm e} = 10^{4}$~K was always assumed althought this is a lower limit as one effect of shocks is to increase the gas temperature above that value. 
     The emission line flux ratio between the [\ion{Ar}{iv}] lines at 4711~\AA~ and 4740~\AA\ is a suitable gas density tracer \citep{wang04}. We observed these two lines in the nuclear and circum-nuclear region. However, because they are weak features, we summed up the three available spectra to increase the S/N. The resulting line flux ratio  4711~\AA/4740~\AA\ is 0.83$\pm$0.17, which points to $n_0$  in the range 4.2$-13\times10^3$cm$^{-3}$. Notice that the ionization energy of the Ar$^{+3}$ ion is 40.74~eV. Therefore, this ion should be more representative of the density of the CL gas than S$^+$ (10.36 eV), for which a $n_0 \sim 10^3$cm$^{-3}$ was measured.
     	
	 All CLOUDY models with $n_{\rm 0} \geq ~10^3$~cm$^{-3}$ produce a very compact coronal emission region, of radius smaller than 50~pc. Indeed, emission regions larger than 100~pc, as observed for [\ion{Si}{vi}] and [\ion{Si}{vii}], are only possible for densities $\leq$100~cm$^{-3}$. For this reason, we discard that this mechanism alone could dominate the production of the coronal lines. \citet{ROA17} tested the scenario where the AGN luminosity  is higher than the one currently estimated here by a factor 50 times larger ($L_{\rm bol}=1.26\times10^{44}$ erg~s$^{-1}$) while keeping the remaining parameters fixed. This explores the possibility of any potential AGN variability, implying that the NLR gas ionization occured when the source was in a higher state. The results     (see Figure 14 of \citealt{ROA17}) show that an AGN with such luminosity is able to sustain extended [\ion{Si}{vi}] and [\ion{Si}{vii}] emission only if $n_{\rm e} \leq$500~cm$^{-3}$, supporting our hypothesis that mechanical heating in ESO~428-G14 must be also considered to explain its large extended high-ionization region.
		
	 Models of \citet{Contini01} are useful to test the effects of shocks coupled to photoionization by the central source. The ionization due to both the primary radiation (AGN) and the diffuse radiation generated by the free$-$free and free$-$bound transitions of the shocked and photoionized gas, as well as the collisional ionization, are all accounted for. The shock velocity, $V_{\rm s}$, the ionizing flux from the central source at the Lyman limit reaching the cloud, $F_{\rm h}$ (in units of cm$^{-2}$~s$^{-1}$~eV$^{-1}$), the preshock density, $n_0$, and the preshock magnetic field, $B_0$ are the main input parameters.
	
	 Tables 1-12 of \citet{Contini01} show that coronal lines become observable for $V_{\rm s} \geq 200$~km\,s$^{-1}$ and $n_0 \geq$ 200~cm$^{-3}$.  For instance, their models 30, 39, 63 and 76, with $V_{\rm s}$ between 200 and 500~km\,s$^{-1}$ and $n_0$ in the range 200$-$300~cm$^{-3}$ produce [\ion{Si}{vi}]/Br$\gamma$~ratios of up to $\sim$13, accounting for all the measured ratios shown in Table~\ref{tab:linefit}. Considering that the photoionization models alone fail at reproducing the observed flux ratios at the 100 parsec scales, the presence of shocks plus photoionization by the central source turns out to be the most consistent scenario. Moreover, the combination of clouds with preshock densities and shock velocities in the interval above  produces downstream densities, which is what we actually measure in the NLR, in the range 4$\times 10^3 - 1.7\times 10^4~$~cm$^{-3}$.

    \subsection{Energetics of the outflowing material}
    \label{sec:energetics}
    
    \begin{table}
    \begin{center}
    \caption[Outflowmass]
    {Outflow mass and mechanical power for the most important structures of [\ion{Si}{vi}] emission observed in ESO\,428-G14.}
    \tabcolsep=0.11cm
    \begin{tabular}{ccccccc}
    \hline \hline
    Region & l \& w & $v_{\rm out}$ & $\sigma$ & $n_{\rm e}$ & $\dot{M}_{\rm out}^1$ & $\dot{E}_{\rm kin}^2$ \\
     &   (pc)  & (km s$^{-1}$) & (km s$^{-1}$) & (cm$^{-3}$) & & \\
    \hline 
    b1 & 21;15 &	250	&150 & 4350	 & 3.5 & 9.5 \\
    b2 & 19;14 &	200	& 334 &	4350 & 2.4 & 11.3 \\ 
    b3 & 50:34 &	100	& 150 & 4350  & 7.8 & 8.0 \\ 
    b4 & 35;19 &    100 & 80 & 4350  & 3.0 & 1.6 \\ 
    \hline
    \end{tabular}
    \begin{minipage}{8cm}
    $^1$ In units of $M_\odot$\, yr$^{-1}$; ~  $^2$Units of  $\times10^{40}$~erg\,s$^{-1}$
    \end{minipage}
    \label{tab:energetics}
    \end{center}
    \end{table}
    
    We use the velocity and the resolved morphology of the outflowing coronal gas  associated to the broad components detected in [Si\,{\sc vi}] (Fig.~\ref{fig:linefit}) to derive the mass outflow rate and the mechanical power injected in the ISM. Assuming ellipsoidal morphologies for the bubbles b1 to b4, corresponding outflow velocities $v_{\rm out}$ (col.~3 of Table~\ref{tab:energetics}) and velocity dispersions $\sigma$ (col.~4), the mass outflow rate (in units of M$_\odot$\, yr$^{-1}$) and the mechanical power (in units of erg\,s$^{-1}$) can be derived in the following way:

\begin{equation} \label{eq:mass}
\begin{aligned}
\dot{M}_{\rm out} = \frac{4}{3} \pi m_{\rm p} ~ n_{\rm e} ~ l~ w ~ f~v_{\rm out}; & & \dot{E}_{\rm kin} = &  \frac{1}{2}\dot{M}_{\rm out} (v^2_{\rm out} + \sigma^2)
\end{aligned} 
\end{equation}

\noindent where $m_{\rm p}$ is the proton mass,  $n_{\rm e}$ is the electronic density, set to $4.35\times 10^3$~cm$^{-3}$ and $l$ and $w$ refer to the length and the width radius of each blob, respectively. The filling factor, $f$, was set to 0.1. The resolved morphology of the coronal gas, of tens of parsecs scale, points out to this number as a lower limit to $f$. Indeed, theoretical and observational arguments presented by \citet{sharp10} showed that for ionized gas winds in nearby AGN and starburst galaxies, $f$ should vary over the range 0.1$-$1.

    From the values of Cols.~2 to ~5 of Table~\ref{tab:energetics} we found that  the outflowing mass is in the range 3-8~M$\odot$~yr$^{-1}$. 
    The corresponding kinetic power varies between 1-6\% of the total radiative output of the AGN. If we consider the range of densities suggested by the Ar$^{+3}$ ratio (2410-7640~cm$^{-3}$), the outflowing mass is between 1.7 and 14~M$\odot$~yr$^{-1}$, with the kinetic power varying between 0.4 and 10\%, respectively.
    Note that the figures above still carry an uncertainty by a factor of 2 if we take into account that the broad component, on average, represents 40\% of the total [\ion{Si}{vi}] flux and that the velocities have not been corrected for projection effects. However, for a low-accreting source that has a low-power radio jet (log$P_{\rm 1.4 GHz} = 21.81$ W\,s$^{-1}$, \citealt{Ulvestad89}), two orders of magnitude below that found in powerful radio-loud sources \citep{Morganti05}, the numbers above allow us to conclude that LLAGNs are fully capable of producing energetic outflows that may significantly impact the ISM in the inner hundreds of parsecs of the central engine. We note that our technique calculates masses and outflow rates purely from the geometry of the system, independent of the observed emission line luminosity, and yields an increase in mass with density. In approaches based on the \ion{H}{i} lines, the mass will decrease with increasing density for a fixed emission line luminosity but is heavily dependent on extinction and underlying stellar population corrections. Nonetheless, this method may provide a proxy for the total mass in the outflow, calculated as $3\times 10^5$\,M$_\odot$.
			
    \section{Conclusions}
     This work shows the relevance of the kinetic channel as a major way of releasing nuclear energy to the ISM of a LLAGN, up to several percent of the bolometric luminosity, even when driven by radiatively poor radio jets. ESO\,428-G14 is a showcase of this scenario. The derived mass outflow rate, of up to 8~M$\odot$\,yr$^{-1}$, is similar to that found in powerful radio-loud AGN. 
    
    \section*{Acknowledgments}

    We thank the referee for the valuable comments, the funding from FAPESP (D.M), CNPq (D.M and A.R.A), CAPES and CONICYT PIA ACT172033 (Y.D.) and M.A.P. and J.A.F.O. acknowledges the financial support from grant MEC-AYA2015-53753-P. We thank to Marcella Contini for her shock-photionization model calculations.

    \bibliographystyle{mnras}
    \bibliography{ref}

    \label{lastpage}

    \end{document}